\documentclass[conference]{IEEEtran}
\IEEEoverridecommandlockouts
\usepackage{cite}
\usepackage{amsmath, amssymb, amsfonts, amsthm}
\usepackage{algorithm}
\usepackage[noend]{algpseudocode} 
\usepackage{graphicx}
\usepackage{textcomp}
\usepackage{xcolor}
\RequirePackage{xspace}
\usepackage[colorlinks=true, allcolors=blue]{hyperref}
\algrenewcommand\algorithmicrequire{\textbf{Input:}}
\algrenewcommand\algorithmicensure{\textbf{Output:}}

\theoremstyle{plain}

\newtheorem*{theorem*}{Theorem}
\newtheorem*{lemma*}{Lemma}
\newtheorem*{proposition*}{Proposition}
\newtheorem*{conjecture*}{Conjecture}
\newtheorem{fact*}{Fact}

\theoremstyle{definition}

\newtheorem*{definition*}{Definition}
\newtheorem*{question*}{Question}
\newtheorem*{example*}{Example}
\newtheorem*{remark*}{Remark}
\newtheorem*{remarks*}{Remarks}
\newtheorem*{exercise*}{Exercise}
\newtheorem*{assumption*}{Assumption}

\newcommand{\Z}{\mathbb{Z}}

\newcommand{\approxd}{\stackrel{\rm d}{\approx}}
\newcommand{\approxst}{\approx_{\rm st}}

\newcommand{\weq}{\ = \ }

\newcommand{\wapprox}{\ \approx \ }

\newcommand{\E}{\mathbb{E}}
\newcommand{\pr}{\mathbb{P}}

\newcommand{\Barabasi}{Barab{\'a}si\xspace}

\newcommand{\argmin}[1]{\underset{#1}{\arg\,\min}\,}

\def\BibTeX{{\rm B\kern-.05em{\sc i\kern-.025em b}\kern-.08em
    T\kern-.1667em\lower.7ex\hbox{E}\kern-.125emX}}
\begin{document}

\title{Analysis of large sparse graphs using regular decomposition of graph distance matrices}

\author{
\IEEEauthorblockN{1\textsuperscript{st} Hannu Reittu}
\IEEEauthorblockA{\textit{ Big data industrial applications } \\
\textit{VTT Technical Research Centre of Finland}\\
P.O. Box 100, FI-02044 VTT, Finland \\
hannu.reittu@vtt.fi}\\
\IEEEauthorblockN{3\textsuperscript{3rd} Tomi R\"aty}
\IEEEauthorblockA{\textit{ Big data industrial applications } \\
\textit{VTT Technical Research Centre of Finland}\\
P.O. Box 100, FI-02044 VTT, Finland\\
tomi.raty@vtt.fi}\\
\and
\IEEEauthorblockN{2\textsuperscript{nd} Lasse Leskel\"a}
\IEEEauthorblockA{\textit{Dept.~Mathematics and Systems Analysis} \\
\textit{Aalto University School of Science}\\
Otakaari 1, 02150 Espoo,  Finland\\
lasse.leskela@aalto.fi}\\
\IEEEauthorblockN{4\textsuperscript{th} Marco Fiorucci}
\IEEEauthorblockA{\textit{Dep. of Environmental Sciences, Informatics and Statistics} \\
\textit{Ca' Foscari University of Venice}\\
Via Torino 155, 30172 Venezia Mestre, Italy \\
marco.fiorucci@unive.it}
}

\maketitle

\begin{abstract}
Statistical analysis of large and sparse graphs is a challenging problem in data science due to the high dimensionality and nonlinearity of the problem. This paper presents a fast and scalable algorithm for partitioning such graphs into disjoint groups based on observed graph distances from a set of reference nodes. The resulting partition provides a low-dimensional approximation of the full distance matrix which helps to reveal global structural properties of the graph using only small samples of the distance matrix. The presented algorithm is inspired by the information-theoretic minimum description principle. We investigate the performance of this algorithm for selected real data sets and for synthetic graph data sets generated using stochastic block models and power-law random graphs, together with analytical considerations for sparse stochastic block models with bounded average degrees.






\end{abstract}


\section{Introduction}

Graphs are a useful abstraction of data sets with pairwise relations. In case of very large graph data sets, extracting structural properties and performing basic data processing tasks may be computationally infeasible using raw data stored as link lists, whereas the adjacency matrix may be too large to be stored in central memory.
One obvious problem with sampling is the sparsity. The sparsity means that if we pick up two nodes at random, we usually observe no relation between them and it is impossible to create any meaningful low-dimensional model of the network. This prevents using uniform sampling as a tool for learning large and sparse graph structures.

Considerable progress towards statistical inference of sparse graphs has recently been achieved, cf.~\cite{fox,borgs6} and references therein.
Most of these methods are based on counting cycles and paths in the observed graph, possibly with some added randomness to split data and reduce noise sensitivity.
%
%
%
Instead of cycle and path counts, here we suggest an alternative approach based on observed graph distances from a set of reference nodes to a set of target nodes. Such distances form a dense matrix. Of course, also in this case it may not be possible to have a complete matrix for very large networks.  What is required is that for any given pair of nodes belonging to a sample, it is possible to have a relatively good estimate of distance between nodes. This is also a nontrivial task requiring an efficient solution, see e.g. \cite{suel}. In recent experiments good estimates for distance matrix were reported for a billion-node graphs \cite{qi}. Our sampling based approach only requires a sparse sample of the full distance matrix. When the number of reference nodes is bounded, the overall computational complexity of the proposed algorithm is linear in number of target nodes. 

We discuss two different sampling schemes of the reference nodes. The first is uniform sampling, which is a feasible method for graphs with light-tailed degree distributions such as those generated by stochastic block models. The second sampling scheme is nonuniform and biased towards sampling nodes with high betweenness centrality, designed to be suitable for scale-free graphs with heavy-tailed degree distributions.

A crucial step is to obtain a low-rank approximation of the distance matrix based on its sample. For this we suggest to use a suitable variant of the regular decomposition (RD) method developed in \cite{reittuetall,nepusz2008,pehkonenreittu,reittujoensuu,reittubazsonorros, reittuetalljournal}. RD can be used for dense graphs and matrices and it shows good scalability and tolerance to noise and missing data. Because the observed distance matrix is dense, RD should work. The method permutes the rows of the matrix into few groups so that each column within each group of the matrix is close to a constant. We call such row groups \emph{regular}. The regular groups form a partition of the node set. Each group of the partition induces a subgraph, and together these subgraphs form a decomposition of the graph into subgraphs and connectivity patterns between them. This decomposition is the main output of the method. The hypothesis of this paper is that the graph decomposition reveals structure of the sparse and large graphs. For instance, it should reveal small but dense subgraphs in sparse graphs as well as sets of similar nodes that form communities.     

As a theoretical latent model we consider stochastic block models (SBM). SBM is an important paradigm in network research, see e.g.\cite{abbe}. Usually SBM revolves around the concept of communities that are well connected subgraphs with only few links between the communities. We also look for other types of structures different from the community structure, see also \cite{newmanpeixoto}. For instance, in case of web graphs, Internet, peer-to-peer networks etc., we would expect quite different structure characterized, say, by a power-law degree distribution and hierarchy of subnetworks forming tiers that are used in routing messages in the network. The proposed distance based structuring might give valuable information about the large scale structure of many real-life networks and scale into enormous network sizes.

Our approach is stimulated by Szemer\'edi's regularity lemma \cite{Szeme76} which indicates that large dense graphs have decomposition of nodes into a bounded number of groups where most of the pairs are almost like random bipartite graphs. The structure encoded by the regularity lemma ignores all intra-group relations. In our regular decomposition both of these aspects are used and both inter-group and intra-group relations matter.

As a benchmark we consider the famous planted bipartition model \cite{decelle}. It is a random graph and a special case of SBM. As ground truth, there are two communities of nodes with equal number of nodes in each and with two parameters. First parameter is the link probability between nodes inside each community and the second one, the link probability between nodes in different communities. The links are drawn randomly and independently from each other. For such a model, it is known that there is a sharp transition, or 'critical point', of detectability of such a structure depending on the difference between the two parameters \cite{decelle, sly}. The critical point is also located in the area of very sparse graphs, when expected degree is bounded. This example is suitable for testing our method because: having a ground truth, graph sparsity, bounded average degree and the proven sharp threshold. The preliminary results we report here, are promising. It seems that our algorithm is effective right up to the threshold in the limit of large scale. Moreover simulations indicate that such a structure can be found from very sparse and bounded size samples of the distance matrix.   

Besides this benchmark, we demonstrate our method using real-life sparse networks. The first example is a Gnutella peer-to-peer file sharing network, and the second is an undirected Internet's autonomous system network. Both of them heavy-tailed degree distributions \cite{remco}. These graphs are not enormous. However, we treat them as if they were very large. Meaning that they are analyzed by using only a small fraction of the full information in the distance matrix. The computations were run in few nodes of a HPC cluster. Using this facility with $2000$ cores and $40$ terabytes of memory, it is possible to run experiments with much bigger data sets in the near future.


\section{Regular decomposition algorithm}

\subsection{Communities and partition matrices}
Consider a connected (finite, undirected) graph\footnote{Or strongly connected directed graph in a directed setting.} $G$. If the original graph is not connected, we can first do a rough partitioning using the connected components. Here we assume that this simple task has already been carried out. Our goal is to partition a subset $V$ of $n$ nodes of the graph into $k$ disjoint nonempty sets called \emph{communities}. Such a partition can be represented as an ordered list $(Z_1,\dots, Z_{n})$ where $Z_i \in [k]$ indicates the community of the $i$-th node in $V$. For convenience, we will also use an alternative representation of the partition as an $n$-by-$k$ matrix with entries
\[
 R_{iu}
 \weq
 \begin{cases}
  1 &\quad \text{if the $i$-th node of $V$ is in community $u$}, \\
  0 &\quad \text{else}.
 \end{cases}
\]
Such a matrix has binary entries, unit rows sums, and nonzero columns, and will be here called a \emph{partition matrix}.

\subsection{Statistical model for the distance matrix}
\label{sec:Poisson}
The partitioning algorithm presented here is based on observed distances from a set of $m$ reference nodes to a (possibly overlapping) set of $n$ target nodes. Let $D_{ij}$ be the length of the shortest path from the $i$-th reference node to the $j$-th target node in the graph. The target is to find such a partition of nodes that distances from any particular reference node $i$ to nodes in community $u$ are approximately similar, with minimal stochastic fluctuations. This modeling assumption can be quantified in terms of an $m$-by-$k$ matrix $(\Lambda_{iu})$ with nonnegative integer entries representing the average distance from the $i$-th reference node to nodes in community $u$. A simple model of a distance matrix in this setting is to assume that all distances are stochastically independent random integers such that the distance from the $i$-th reference node in to a node in community $u$ follows a Poisson distribution with mean $\Lambda_{iu}$. This statistical model is parametrized by the $m$-by-$k$ average distance matrix $\Lambda$ and the $n$-by-$k$ partition matrix $R$, and corresponds to the discrete probability density function\footnote{We could omit terms with $i=j$ from the product because of course $D_{ii}=0$, but this does not make a big difference for large graphs.}
\[
 f_{\Lambda,R}(D)
 \weq \prod_{i=1}^{m} \prod_{j=1}^{n} e^{-\Lambda_{i Z_j}} \frac{\Lambda_{i Z_j}^{D_{ij}}}{D_{ij}!},
 \quad D \in \Z_+^{m \times n},
\]
having logarithm
\begin{align*}
 \log f_{\Lambda,R}(D)
 &\weq \sum_{i=1}^m \sum_{j=1}^n \sum_{v=1}^k R_{jv} \left( D_{ij} \log \Lambda_{iv} - \Lambda_{iv} \right) \\
 &\quad\quad - \sum_{i=1}^m \sum_{j=1}^n \log(D_{ij}!) .
\end{align*}

Such modeling assumption does not assume any particular distribution of distance matrix, question is about approximating the given distance matrix with a random matrix with parameters that give the best fitting. Such  particular models are used because it results in a simple program, as we see in Algorithm 1. We have also tested it in our previous works with various data, showing good practical performance, \cite{pehkonenreittu,reittujoensuu,pirkko}. 

Having observed a distance matrix $D$, standard maximum likelihood estimation looks for $\Lambda$ and $R$ such that the above formula is maximized. For any fixed $R$, maximizing with respect to the continuous parameters $\Lambda_{iv}$ is easy. Differentiation shows that the map $\Lambda \mapsto \log f_{\Lambda,R}(D)$ is concave and attains its unique maximum at $\hat \Lambda = \hat \Lambda(R)$ where
\begin{equation}
 \label{eq:MLELambda2}
 \hat \Lambda_{iv}(R)
 \weq \frac{\sum_{j=1}^n D_{ij}R_{jv}}{\sum_{j=1}^n R_{jv}}
\end{equation}
is the observed average distance from the $i$-th reference to nodes in community~$v$. As a consequence, a maximum likelihood estimate of $(\Lambda, R)$ is obtained by minimizing the function
\begin{equation}
 \label{eq:NegLogLikelihood}
 L(R)
 \weq
 \sum_{i=1}^m \sum_{j=1}^n \sum_{v=1}^k R_{jv} \left( \hat \Lambda_{iv}(R) - D_{ij} \log \hat \Lambda_{iv}(R) \right) 
\end{equation}
subject to $R \in \{0,1\}^{n \times k}$ having unit row sums and nonzero column sums, where $\hat \Lambda_{iv}(R)$ is given by \eqref{eq:MLELambda2}.

\subsection{Recursive algorithm}

Minimizing \eqref{eq:NegLogLikelihood} is a nonlinear discrete optimization problem with an exponentially large input space of order $\Theta(k^n)$. Hence an exhaustive search is not computationally feasible. The objective function can alternatively be written as $L(R) = \sum_{j=1}^n \ell_{j Z_j}(R)$, where
\begin{equation}
 \label{eq:LikelihoodComp}
 \ell_{jv}(R)
 \weq \sum_{i=1}^m \left( \hat \Lambda_{iv}(R) - D_{ij} \log \hat \Lambda_{iv}(R) \right).
\end{equation}
This suggests a way to find local maximum by selecting a starting value $R^0$ for $R$ at random, and greedily updating the rows of $R$ one by one as long as the value of the objective function decreases.  A local update rule for $R$ is achieved by a mapping $\Phi: \{0,1\}^{n \times k} \to \{0,1\}^{n \times k}$ defined by $
 \Phi(R)_{jv} = \delta_{Z^*_j v}$ where
\begin{equation}
 \label{eq:MStep}
 Z^*_j \weq \argmin{v \in [k]} \ell_{jv}(R).
\end{equation}
Algorithm~\ref{alg:RD} describes a way to implement this method.  This is in spirit of the EM algorithm where the averaging step corresponds to an E-step and the optimization step to an M-step. The algorithm iterates these steps by starting from a random initial partition matrix $R^0$, and recursively computing $R^{t} = \Phi(R^{t-1})$ for $t \ge 1$. The runtime of the local update is $O(km+kn)$, so that as long as the number of communities $k$ and the parameters $s_{\rm max}, t_{\rm max}$ are bounded, the algorithm finishes in linear time with respect to $m$ and $n$ and is hence well scalable for very large graphs. The output of Algorithm~\ref{alg:RD} is a local optimum.  To approximate a global optimum, parameter $s_{\rm max}$ should be chosen as large as possible, within computational resources.

Finally, we describe how the rest of nodes are classified into $k$ groups or communities, after the optimal partition $R^*$ for a given target group and reference group is found. 
Let $i$ denote a node out of original target group. First we must obtain distances of this node to all reference nodes 
$$
(D_{i,j})_{1\leq j\leq m}
$$
Then the node i is classified into group number $\alpha$ according to
$$
\alpha= \arg\min_{1\leq \beta\leq k}\sum_{j=1}^m \left( \hat \Lambda_{j\beta}(R) - D_{ji} \log \hat \Lambda_{j\beta}(R^*) \right). 
$$
The time complexity of this task is dominated by the computations of distances of all nodes to the reference nodes, because for bounded $m$ the above optimization is done in a constant time. According to Dijkstra-algorithm computation of distances from all $N$ nodes to the target nodes takes  $mO(|E|+ N\log N)$. In a sparse graph, that we assume, $|E|\sim N$. Thus, if $m$ is bounded, the overall time complexity is $O(N\log N)$, which is only slightly over the best possible $O(N)$, which is needed just to enlist a partition. This is because the classification phase takes only $O(N)$ time for all nodes.
\begin{algorithm}
    \caption{\label{alg:RD} Regular decomposition algorithm}
    \begin{algorithmic}[1]
    \Statex
    \Require {Distance matrix $D \in \Z_+^{m \times n}$, integers $k, s_{\rm max}, t_{\rm max}$ }
    \Ensure {Partition matrix $R^* \in \{0,1\}^{n \times k}$}
    \Statex
    \Function{RegularDecomposition}{$D, k, s_{\rm max}, t_{\rm max}$}
    \State $L_{\rm min} \leftarrow \infty$
    \For{$s$ in $1,\dots, s_{\rm max}$}
       \State $R \leftarrow$random $n$-by-$k$ partition matrix
           \For{$t$ in $1,\dots, t_{\rm max}$}
           \State $R \leftarrow$ \Call{LocalUpdate}{$R,D$}
       \EndFor
       \State $L \leftarrow L(R)$ given by equation \eqref{eq:NegLogLikelihood}
       \If{$L < L_{\rm min}$}
          \State $R^* \leftarrow R$
          \State $L_{\rm min} \leftarrow L$
       \EndIf
    \EndFor
    \State \Return $R^*$    
    \EndFunction
    \Statex
    \Function{LocalUpdate}{$R, D$}
    \Statex \textit{Averaging step}
    \For{$v$ in $1,\dots,k$}
      \State $n_v \leftarrow \sum_{j=1}^n R_{jv}$
      \For{$i$ in $1,\dots,m$}
         \State $\hat\Lambda_{iv} \leftarrow \sum_{j=1}^n D_{ij}R_{jv}/n_v$
      \EndFor      
      \For{$j$ in $1,\dots,n$}
         \State $\ell_{jv} \leftarrow \sum_{i=1}^m( \hat \Lambda_{iv} - D_{ij} \log \hat \Lambda_{iv})$
      \EndFor
    \EndFor
    \Statex \textit{Optimization step}
    \For{$j$ in $1,\dots,n$}
       \State $Z^*_j \leftarrow \argmin{v \in [k]} \ell_{jv}$
       \For{$v$ in $1,\dots,k$}
       \State $R^*_{jv} \leftarrow 1(Z^*_j = v)$
       \EndFor
    \EndFor
    \State \Return $R^*$
    \EndFunction
    \end{algorithmic}
\end{algorithm}

\subsection{Estimating the number of groups}

The regular decomposition algorithm presented in the previous section requires the number of groups $k$ as an input parameter. However, in most real-life situations this parameter is not a priori known and needs to be estimated from the observed data.
%
%
%
%
The problem of estimating the number of groups $k$ can be approached by recasting the maximum likelihood problem in terms of the minimum description length (MDL) principle \cite{Rissanen_1983,grunwald} where the goal is to select a model which allows the minimum coding length for both the data and the model, among a given set of models.
%
When the set of models equasl the Poisson model described in Sec.~\ref{sec:Poisson}, then the $R$-dependent part of the coding length equals the function $L(R)$ given by \eqref{eq:NegLogLikelihood}, and a MDL-optimal partition $R^*$, given $k$, corresponds to the minimal coding length
\[
 R^* = \argmin{R} L(R).
\]
It is not hard to see that $L(R^*)$ is monotonously decreasing as a function of $k$, and in MDL a balancing term, the model complexity, is added to select the model that best explains the observed data. However, in all of our experiments, the negative log-likelihood as a function of $k$ becomes essentially a constant above some value $k^*$.
Such a knee-point $k^*$ is used as an estimate for the number of groups in this paper.
Thus we are using a very simplified version of MDL, since it was found sufficient in our cases of examples. In a more accurate analysis one should use model complexity in higher detail. Some early work towards this direction includes \cite{reittubazsonorros}. 

%


\section{Theoretical considerations}
\subsection{Planted partition model}
\label{sec:SBM}

A stochastic block model (SBM) with $n$ nodes and $k$ communities is a statistical model parametrized by a nonnegative symmetric $k$-by-$k$ matrix $(W_{uv})$ and a $n$-vector $(Z_i)$ with entries in $[k]$. The SBM generates a random graph where each node pair $\{i,j\}$ is linked with probability $W(Z_i,Z_j)$, independently of other node pairs. For simplicity, we restrict the analysis to the special case with $k=2$ communities where the link matrix is of the form
\[
 W
 \weq
 \begin{bmatrix}
  a/n & b/n \\
  b/n & a/n
 \end{bmatrix}
\]
for some constants $a,b > 0$. This model, also known as the \emph{planted partition model}, produces sparse random graphs with link density $\Theta(n^{-1})$, and is a de facto benchmark for testing the performance of community detection algorithms. As usual, we assume that the underlying partition is such that both communities are approximately of equal size, so that the partition matrix $R_{iu} = \delta_{Z_i,u}$ satisfies $\sum_{i=1}^n R_{iu} = (1+o(1)) \frac{n}{2}$. If $a>b$, there are two communities that have larger internal link density than link density between them.  A well-known result states that for $n \gg 1$, partially recovering the partition matrix from an observed adjacency matrix is possible if
\begin{equation}
 \label{eq:KS}
 (a-b)^2 > 2(a+b),
\end{equation}
and impossible if the above inequality is reversed. This result, called Kesten--Stigum threshold, was obtained in semi-rigorous way \cite{decelle} and then proved rigorously in~\cite{sly}. 

\subsection{Expected and realized distances}
Our aim is to have analytical formulas for distances $D_{ij}$ in a large graph generated from SBM. This question was addressed in \cite{bickel} using spectral approach, where limiting average distances were found. We need the next to the leading term of the average distance. Although these calculations are not rigorous, it is well-known that in case of classical random graph similar approach produces a distance estimate that is asymptotically exact. That is why we believe that such an analysis makes sense in case of SBM as well.

To analyze distances, let us first investigate the growth of the neighborhoods from a given node as a function of the graph distance. Let us denote the communities by $V_u = \{i: Z_i=u\}$ for $u=1,2$. Fix a node $i \in V_1$ and denote by $n_u(t)$ the expected number of nodes in community $u$ at distance $t$ from $i$. Note that each node has approximately $a/2$ neighbors in the same community and approximately $b/2$ neighbors in the other community. Moreover, due to sparsity, the graph is locally treelike, and therefore we get the approximations
\begin{align*}
 n_1(t) &\wapprox \frac{1}{2}an_1(t-1)+\frac{1}{2}bn_2(t-1)\\ 
 n_2(t) &\wapprox \frac{1}{2} bn_1(t-1)+\frac{1}{2}an_2(t-1).
\end{align*}
Writing $N(t) = (n_1(t), n_2(t))^T$, this can be expressed in matrix form as
$
 N(t) \approx AN(t-1),
$
where
\[
 A
 = \frac{1}{2}
 \begin{pmatrix}
 a & b \\
 b & a 
 \end{pmatrix}.
\]
As a result, $N(t) \approx A^tN(0)$ with $N(0) = (1, 0)^T$. The matrix $A$ has a pair of orthogonal eigenvectors  with eigenvalues:
\[
 e_1=\frac{1}{\sqrt 2}
 \begin{pmatrix} 
 1 \\
 1
\end{pmatrix},
\quad \lambda_1= \frac{a+b}{2}
\]
and
\[
 e_2=\frac{1}{\sqrt 2}
 \begin{pmatrix} 
 -1 \\
 1
 \end{pmatrix},
 \quad
 \lambda_2 = \frac{a-b}{2}.
\]
According to the spectral theorem, we can diagonalize the matrix $A$ and conclude that its powers are given by
\[
 A^t
 = \lambda_1^t e_1e_1^T+\lambda_2^t e_2e_2^T.
\]
As a result, the expected numbers of nodes of types $u=1,2$ at distance $t$ from a node of type $1$ are approximated by
\[
 N(t)
 \wapprox A^tN(0)
 \weq \frac{1}{2}
 \begin{pmatrix} 
 \lambda_1^t+\lambda_2^t \\
 \lambda_1^t-\lambda_2^t
 \end{pmatrix}.
\]
Moreover, if $m_u(t) = \sum_{s=1}^t n_u(s)$, then
\begin{align*}
 m_1(t)
 &\wapprox
 \frac12\left( -2 + \frac{\lambda_1}{\lambda_1-1} \lambda_1^t + \frac{\lambda_2}{\lambda_2-1} \lambda_2^t \right), \\
 m_2(t)
 &\wapprox
 \frac12\left( -2 + \frac{\lambda_1}{\lambda_1-1} \lambda_1^t - \frac{\lambda_2}{\lambda_2-1} \lambda_2^t \right).
\end{align*}

Next we want to find and estimate for the average distance $d_1$ (resp.\ $d_2$) from a node in $V_1$ to another node in $V_1$ (resp.\ $V_2$). We use the heuristic that the distances from a node to all nodes in the same group are well concentrated and close to each other. Under this assumption, we expect that $d_1$ and $d_2$ approximately solve the equations $m_1(d_1) = n/2$ and $m_2(d_2) = n/2$.
%

%
We get the equations for the distances:  
\begin{eqnarray}
\frac{\lambda_1^{d_1+1}}{\lambda_1-1}+\frac{\lambda_2^{d_1+1}}{\lambda_2-1}-2=n\\\nonumber
\frac{\lambda_1^{d_2+1}}{\lambda_1-1}-\frac{\lambda_2^{d_2+1}}{\lambda_2-1}-2=n.
\end{eqnarray}
We are interested in leading order of difference  of $d_2-d_1$ for $n\rightarrow\infty$. Because $\lambda_1>\lambda_2$ due to $a > b$, and $d_1,d_2 \to \infty$, we can use following iterative solution scheme. For $d_1$, we have:
$$
\lambda_1^{d_1}=\frac{\lambda_1-1}{\lambda_1}n+2\frac{\lambda_1-1}{\lambda_1}-\frac{\lambda_2(\lambda_1-1)}{\lambda_1(\lambda_2-1)}\lambda_2^{d_1}.
$$
as a result, the equation we want to iterate is:
$$
d_1\log\lambda_1=\log\left(\frac{\lambda_1-1}{\lambda_1}n\right)+\log\left(1+\frac{2}{n}- \frac{\lambda_2}{\lambda_2-1}\frac{\lambda_2^{d_1}}{n}\right).
$$
By expanding the second logarithm in series of powers of $1/n$, we get the leading terms of the solution:
$$
d_1\approx\frac{\log\left(\frac{\lambda_1-1}{\lambda_1}n\right)}{\log \lambda_1}-c n^{\alpha-1},
$$
where $c>0$ is a constant and
$$
\alpha= \frac{\log\lambda_2}{\log\lambda_1}, \quad c=\frac{1}{\log \lambda_1}\frac{\lambda_2}{\lambda_2-1}\lambda_2^\beta,\quad \beta= \frac{\log\frac{\lambda_1-1}{\lambda_1}}{\log\lambda_1}.
$$
A similar procedure yields:
$$
d_2\approx\frac{\log\left(\frac{\lambda_1-1}{\lambda_1}n\right)}{\log \lambda_1}+c n^{\alpha-1}.
$$
Because $\alpha < 1$, both $d_1/\log n$ and $d_2/\log n$ have the same limit $1/\log\lambda_1$. 

We conjecture that above the Kesten--Stigum threshold \eqref{eq:KS} the cost function, used in RD to partition graph distance matrix of the giant component of the graph generated from two part SBM, has a deep minimum corresponding to correct partition. More precisely, the cost of misplacing one node from the correct partition grows to infinity as $n\rightarrow\infty$.

First we conjecture that the found distance estimates of $d_1$ and $d_2$ are asymptotically equal to expected distances in a random graph corresponding to the Planted Partition model. For a node $i \in V_1$ and nodes $j \in V_1 \setminus\{i\}$ and $j' \in V_2$,
\[
 \E D_{i,j}
 \approx d-\delta,
 \quad
 \E D_{i,j'} \approx d+\delta,
\]
where $d=\frac{\log\left(\frac{\lambda_1-1}{\lambda_1}n\right)}{\log \lambda_1}$ and $\delta= c n^{\alpha-1}$, corresponding to the approximations in the previous section. We also conjecture that for any $i \in V_1$, we have with high probability,
\begin{align*}
 \sum_{j \in V_1} D_{i,j}
 &\weq \frac{n}{2}(d-\delta)+O(\sqrt n), \quad \\
 \sum_{j \in V_2} D_{i,j} &\weq \frac{n}{2}(d+\delta)+O(\sqrt n),
\end{align*}
which is quite plausible if the first conjecture is true. The error term $O(\sqrt n)$ can be neglected if $\alpha > \frac{1}{2}$, which is equivalent to being above the Kesten--Stigum threshold \eqref{eq:KS}, which we assume from now on. If all nodes of the graph are partitioned correctly, then the cost \eqref{eq:LikelihoodComp} of target node $j$ in community $V_1$ is approximately
\[
 \ell_{j}
 \approx \frac{n}{2}(d+\delta-(d+\delta)\log(d+\delta)+\\d-\delta-(d-\delta)\log(d-\delta)).
\]
If we switch the community $j$ to be $V_2$ then this cost changes into
\[
 \ell'_j
 \approx \frac{n}{2}(d+\delta-(d-\delta)\log(d+\delta)+\\d-\delta-(d+\delta)\log(d-\delta)).
\]
As a result,
\begin{align*}
 \ell'_j - \ell_j
 &\wapprox n\delta\log\left(\frac{d+\delta}{d-\delta}\right) \\
 &\wapprox 2n\frac{\delta^2}{d} \\
 &\weq 2\log(\lambda_1)c^2n n^{2\alpha-2}/\log n\\
 &\weq \frac{2\log(\lambda_1) c^2}{\log n}n^{2\alpha-1}. 
\end{align*}
As a result if $\alpha > \frac12$ or equivalently \eqref{eq:KS}, the difference has infinite limit. This heuristic derivation suggests that the regular decomposition algorithm is capable of reaching the fundamental limit of resolution of the planted bipartition model.

\section{Experiments with simulated data}

\subsection{Planted partition model}

We investigate empirically the performance of the regular decomposition algorithm to synthetic data sets generated by the planted partition model described in Sec.~\ref{sec:SBM}. This is an instance of a very sparse graph with bounded degrees and with only two groups. In this case we argue that uniform random sampling of reference nodes will do. Here it is possible to compute full distance matrix up to sizes of $10000$ nodes and sampling is not necessary.

\begin{figure}
\centering
\includegraphics[width=7cm]{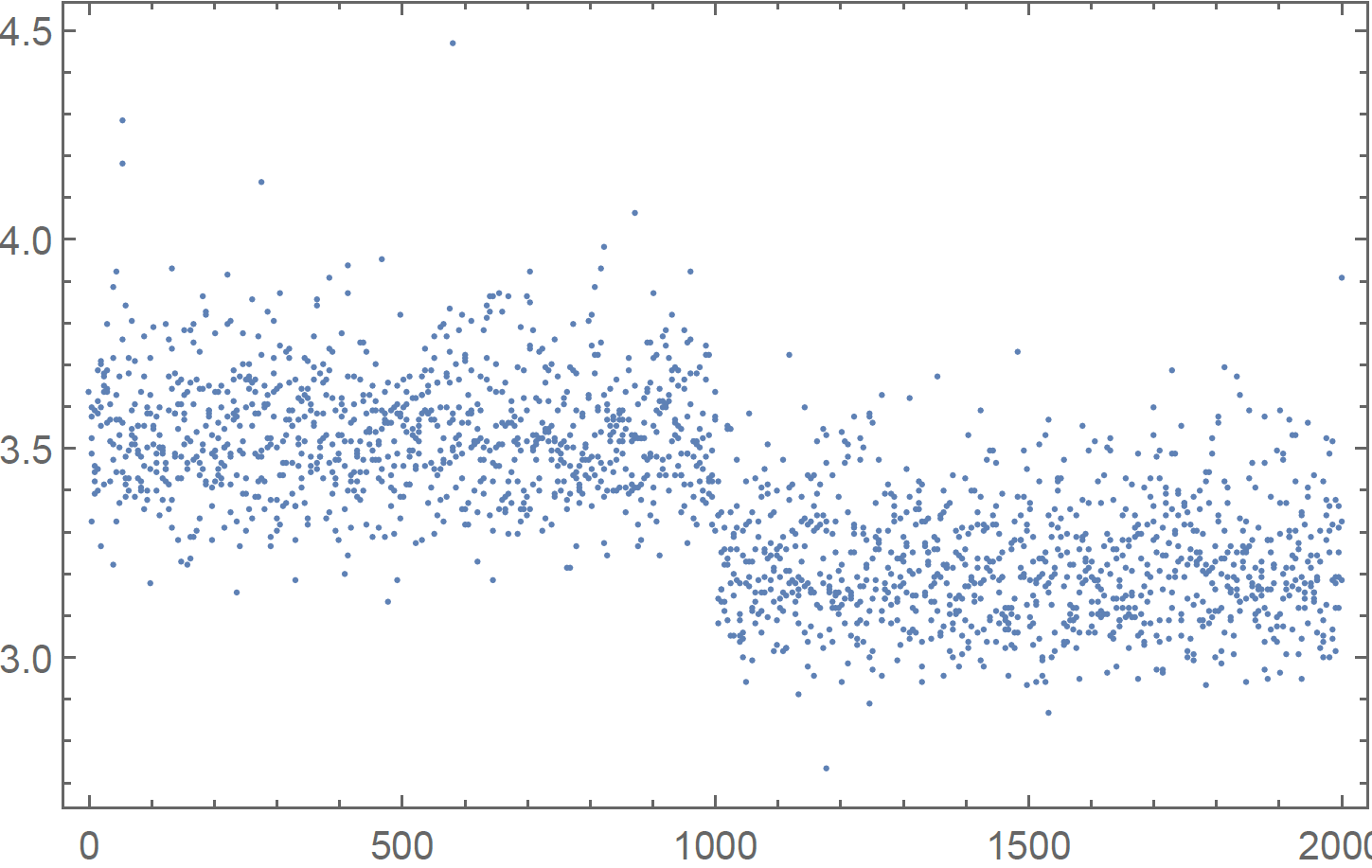}
\caption{Average distance ($y$-axis) $\hat\Lambda_{i2}$ to nodes in group 2 from all nodes $i$ (on $x$-axis) in the planted partition model. The $x$-axis nodes with indexes from $1$ to $2000$. The first $1000$ nodes belong to the first group, and rest to the other one. The right half of the points, correspond to the distances within the same community, the left half of points corresponds to the distances between nodes in different communities. As can be seen, intra-group distances are systematically lower, although with some substantial variance. This is a case of tolerable variance which permits structure to be found.}
\label{noise}
\end{figure}

For our test, we generated a graph with parameters $n=2000$, $a=20$, and $b=2$.  Another similar experiment was done with $10000$ nodes. Next we computed the shortest paths between all pairs of nodes and formed a distance matrix $D$. RD was able to detect the structure with around 1 percent error rate. The average of one regular group shows that the distance has quite high level of noise, see Fig.~\ref{noise}. The reason why the communities become indistinguishable is probably in the increasing level of the variance. Below the threshold it is always too large, no matter how large $n$ is and above the threshold the communities can be detected provided $n$ is large enough. This is the conclusion of experiments not shown in this work.  

\begin{figure}
\centering
\includegraphics[width=6cm]{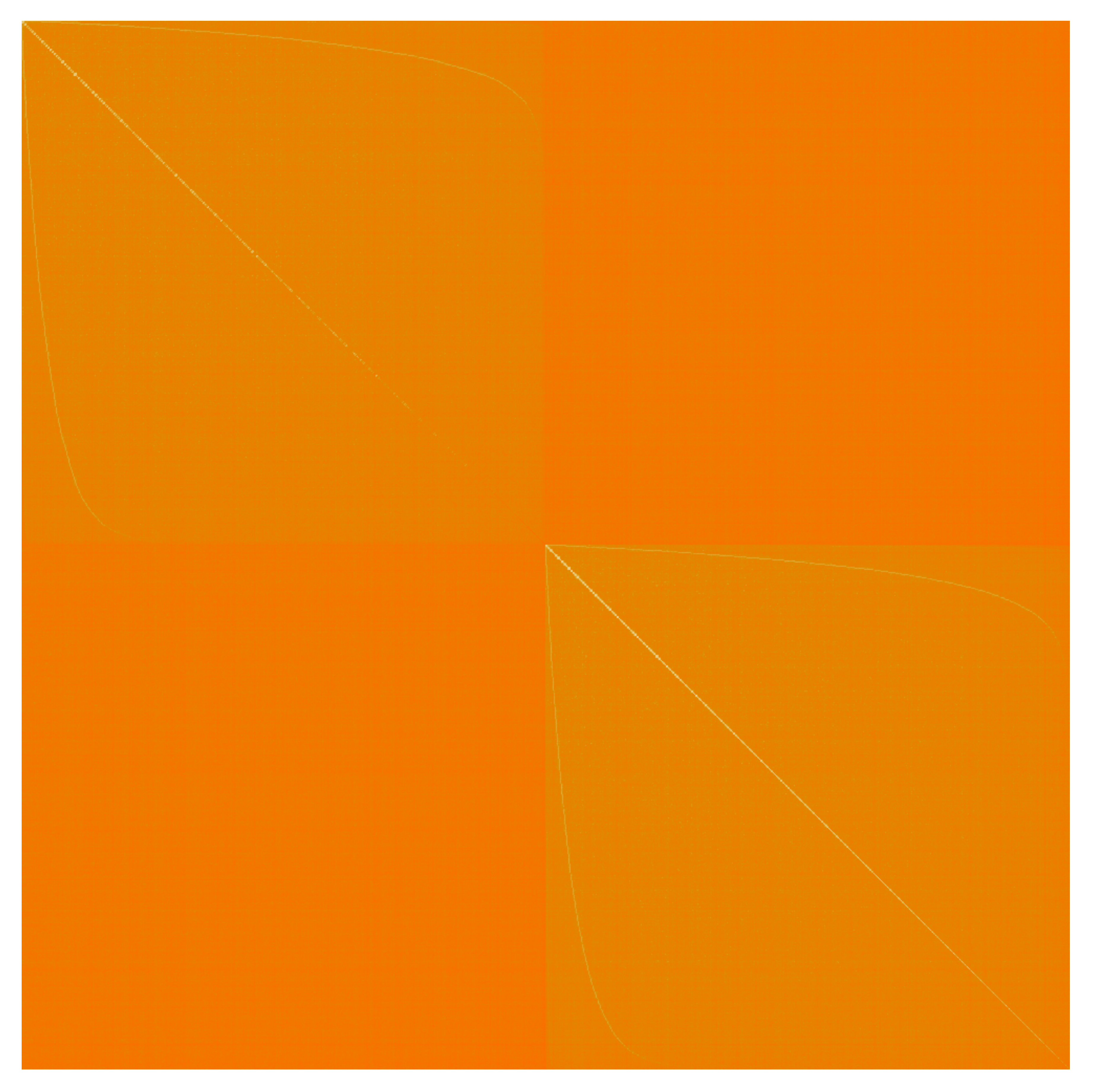}
\caption{Distance matrix of a graph with 10000 nodes sampled from the planted partition model, when nodes are labeled according to two true groups. The intra-block distances (average 4.75) are smaller than the inter-block distances (average 5.0).}
\label{dd}
\end{figure}

Next we did experiments with 10000 nodes. In this particular case it looks that our method works better than a standard community detection algorithm of Girvan-Newman type. See Fig. \ref{dd} for graphical presentation.  

\begin{figure}
\centering
\includegraphics[width=8cm]{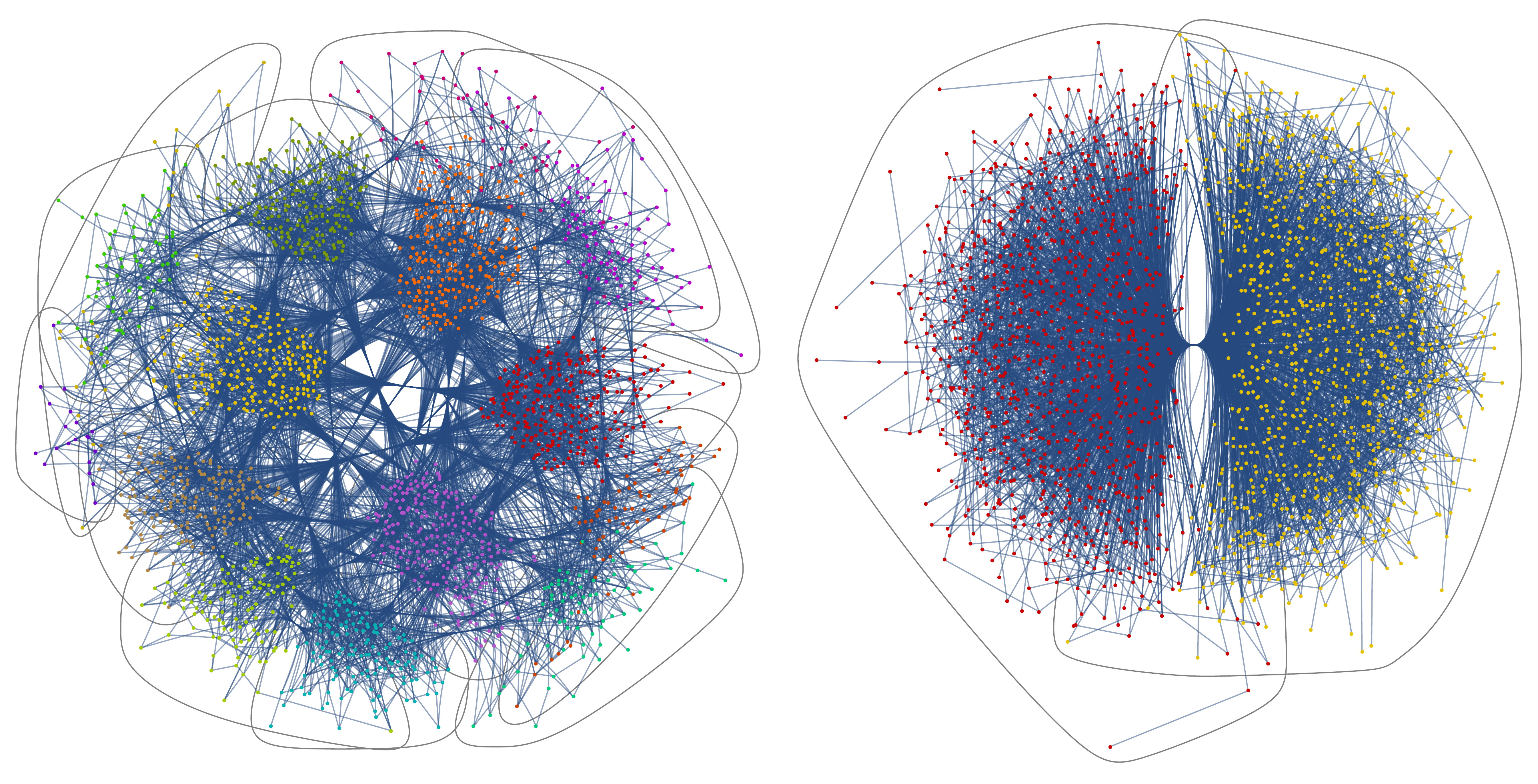}
\caption{Left: the community structure found with the Mathematica's FindGraphCommunities, (that uses, to our knowledge, Girvan-Newman algorithm) applied to our case of planted bipartition graph. It completely fails in detecting the right communities; instead of two correct $15$ communities are found. Right: the almost correct communities found by RD. However, the Mathematica command was not forced to find just two communities. }
\label{community}
\end{figure}

As a sanity check we also used a usual community detection algorithm found in Wolfram Mathematica library. It was not capable of finding the true communities, see Fig. \ref{community}. The RD algorithm using the $D$-matrix, was able to find the communities correctly, with only a handful of misclassified nodes.

\subsection{Sampled distance matrices}

To investigate experimentally how many reference nodes are needed to obtain an accurate partitioning of a set of $n$ target nodes, we sampled a set of $m$ reference nodes uniformly at random, and ran the regular decomposition algorithm on the corresponding $m$-by-$n$ distance matrix.

\begin{figure}
\centering
\includegraphics[width=8cm]{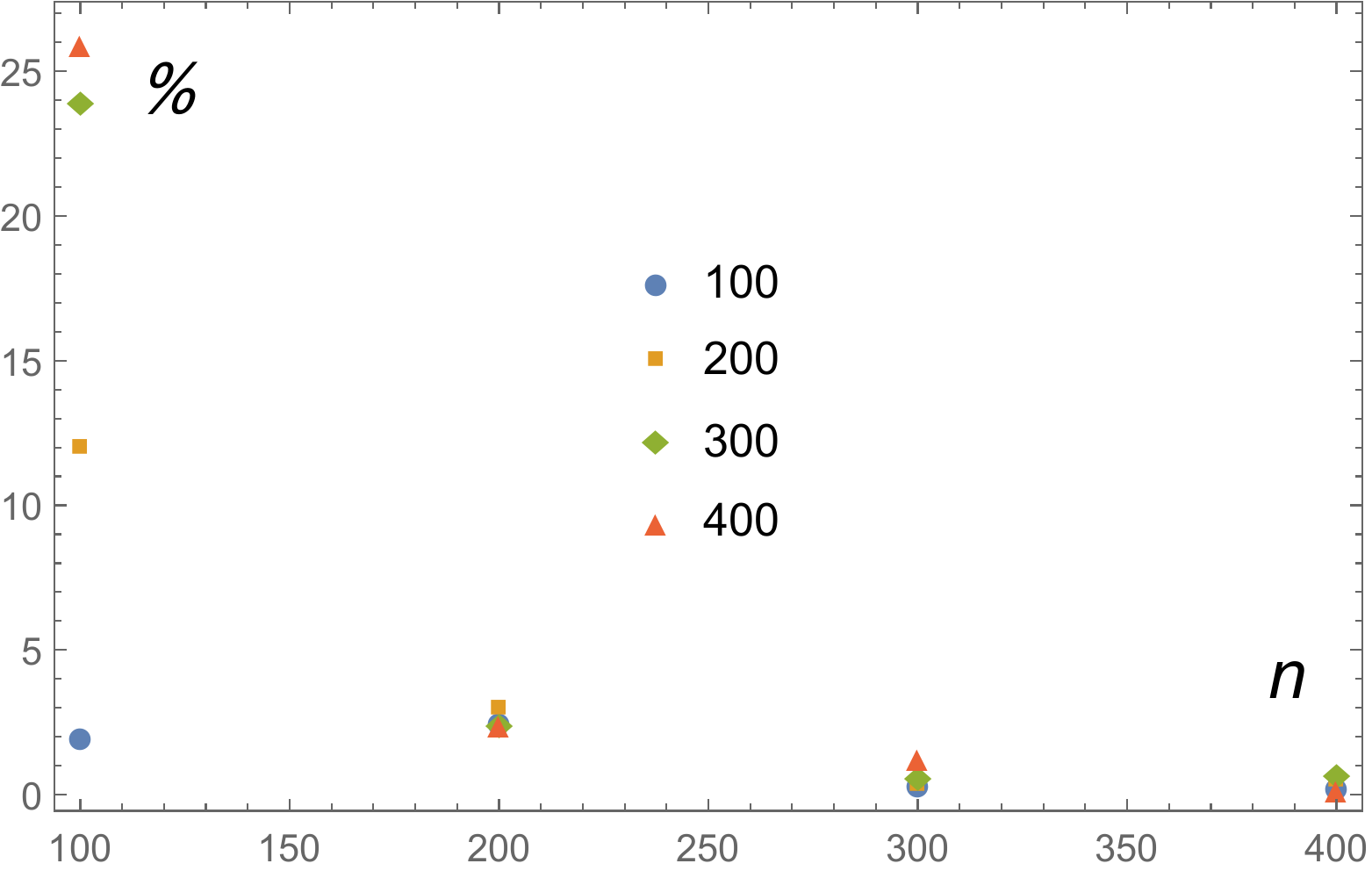}
\caption{Misclassification rates ($y$-axis) for synthetic data generated using a planted bipartition model shown in Fig.~\ref{dd}. The number of (randomly selected) target nodes has values $100,200,300,400$, indicated by colored markers. For each case a random sample of $m$ reference nodes was selected.  The target node set was partitioned into $k=2$ blocks and compared with the ground truth classification.  When $m = 400$, the error rate is less than $1\%$  in all cases.}
\label{succes}
\end{figure}

It appears that even a modest sample of about $m=400$ reference nodes is enough to have almost error free partitioning, see Fig.~\ref{succes}. It appears that with $m \ge 400$ reference nodes, a set of $n=100$ target nodes can be accurately partitioned into $k=2$ communities using RD, with error rate less than $1\%$.  For larger sets of target nodes, the results appear similar. This suggests that such a method could work for very large graphs using this kind of sampling.

The regular decomposition algorithm also produces an estimated $m$-by-$k$ average difference matrix $(\hat\Lambda_{iu})$. This model can be used to classify all nodes in the graph in linear time. To do this, we must compute distances to the $m=400$ reference nodes, compute the negative log-likelihood for two groups based on $(\hat\Lambda_{iu})$, and place the node into the class with a smaller negative log-likelihood. All computations take just a constant time and that is why the linear scaling. 

As a conclusion, we conjecture that for very large and sparse networks the distance matrix RD could be an option to study community structures. 

The RD method seems to have better resolving power than community detection algorithms based on adjacency matrix and could work with sparse samples of data and thus scaling to extremely large networks. For a rather flat topology the uniform sampling method for distance matrix might be sufficient.

\subsection{Preferential attachment models}

The degree distributions of many social, physical, and biological networks have heavy tails resembling a power law. For testing network algorithms and protocols on variable instances of realistic graphs, synthetic random graph models that generate degrees according to power laws have been developed \cite{remco}. 

The main purpose of this exercise is to test sampling approach versus the full analysis.  We used an instance of a preferential attachment model (\Barabasi--Albert  random graph) with $5000$ nodes. The construction starts from a triangle. Then nodes are added one-by-one. Each incoming node makes $3$ random links to existing nodes, and the probability of link is proportional to the degree of a node (preferential attachment). The result is somewhat comparable to the Gnutella network. However, in Gnutella networks instead of hubs, we had some more complicated dense parts. 

\begin{figure}
\centering
\includegraphics[width=8cm]{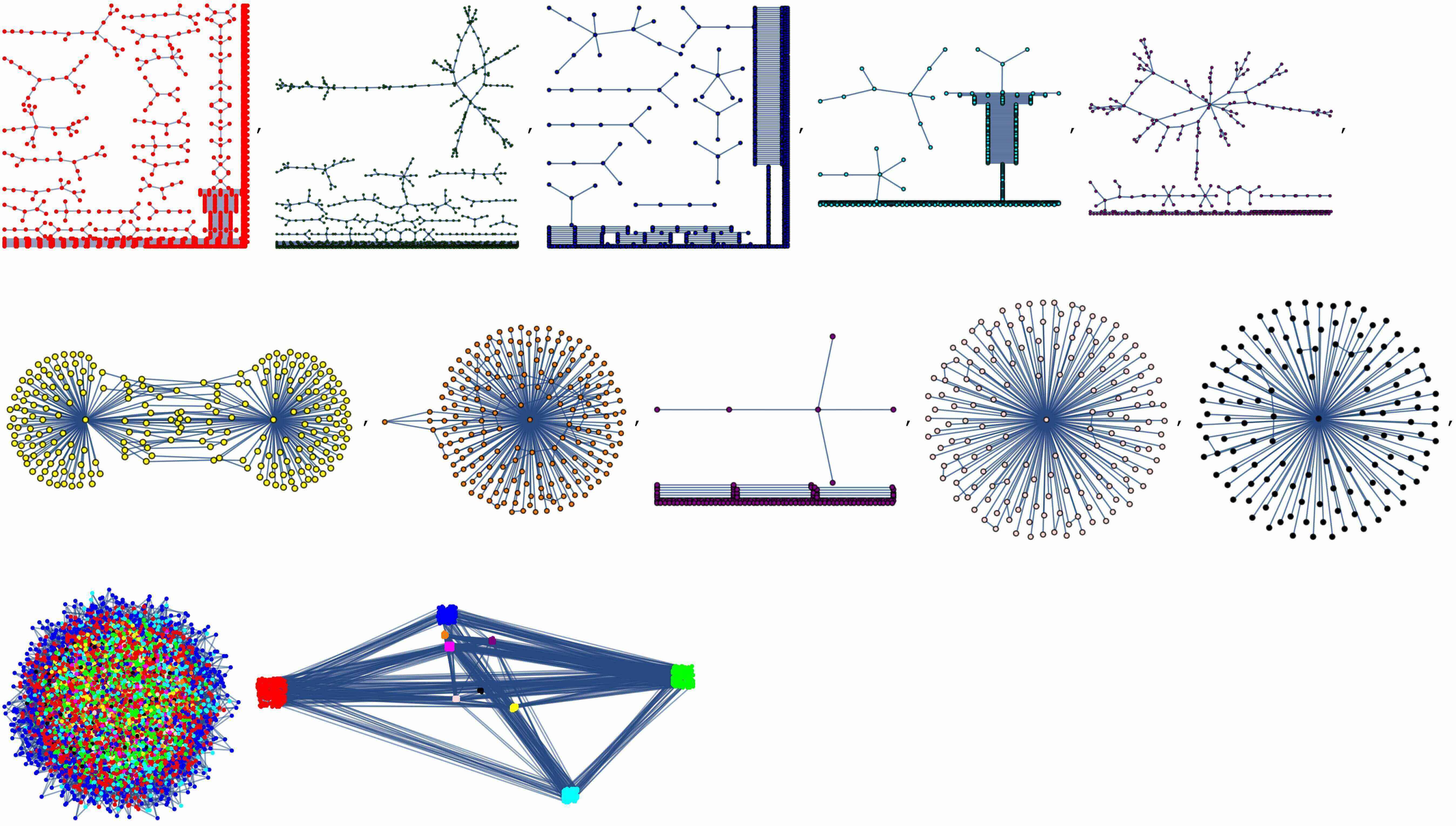}
\caption{At the bottom are the original network and its RD in  $10$ groups. Above are the internal structure of $10$ sub-networks found with RD. Notably there are $4$ groups that are "hubs", having one or two high degree nodes in a star-like topology. In a power-law graph such hubs are known to be essential.  }
\label{barabasi}
\end{figure}

\begin{figure}
\centering
\includegraphics[width=8cm]{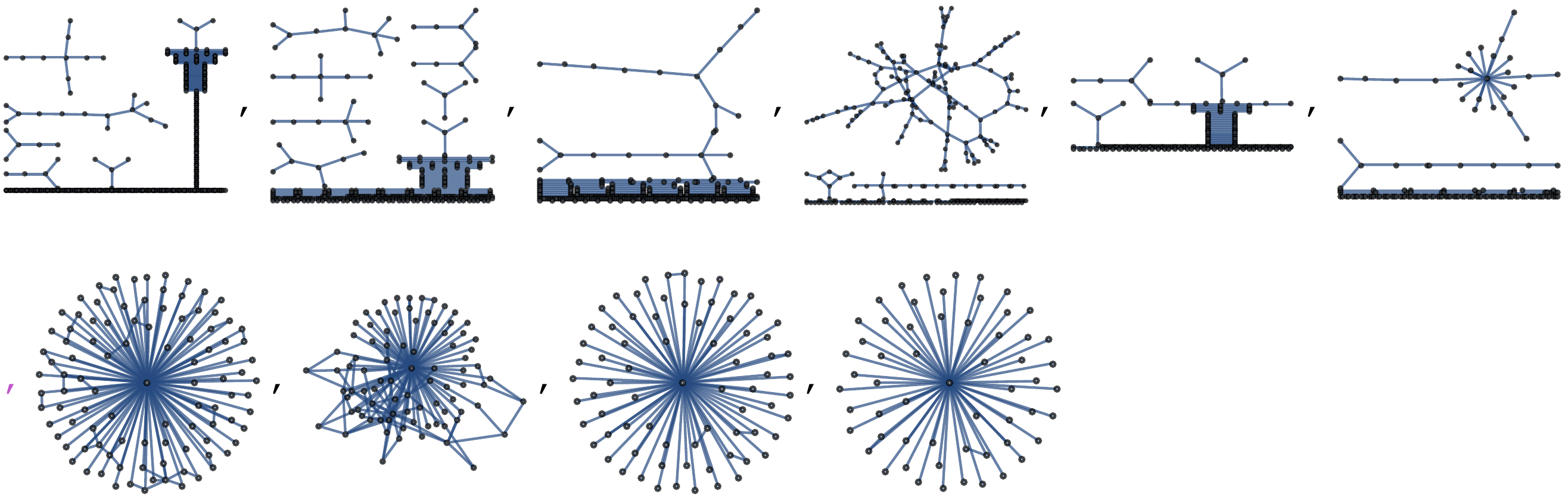}
\caption{Induced subgraphs of a preferential attachment graph corresponding to 10 groups discovered using the regular decomposition algorithm using a small set of $m$ reference nodes of high betweenness centrality. Although the found groups are not identical to the ones found with the full distance matrix ($m=n$), they are correlated to them, and the hub-like subnetworks are also found.}
\label{barabasigroups}
\end{figure}

To achieve scalability, instead of using full distance information between all $n=5000$ nodes, we wish to restrict to distances to a small number of reference nodes. A main problem with the sampling of reference nodes is that high-degree core nodes are unlikely to show up in uniformly random samples. This is why decided to investigate the following nonuniform sampling scheme. The set of reference nodes was generated as the set of $m \approx 1000$ nodes which appeared in shortest paths between a randomly chosen set of 100 pairs of nodes. Distances to such reference nodes of high betweenness centrality are a strong indicator about distances between any two nodes because most short paths traverse through the central nodes. Next we ran the regular decomposition algorithm with $m$ reference nodes to partition the set of $n$ target nodes into $k=10$ blocks. We get a quite similar result as the one for the entire distance matrix, see Fig~\ref{barabasigroups}.


\section{Experiments with real data}
\subsection{Gnutella network}

\begin{figure}
\centering
\includegraphics[width=8cm]{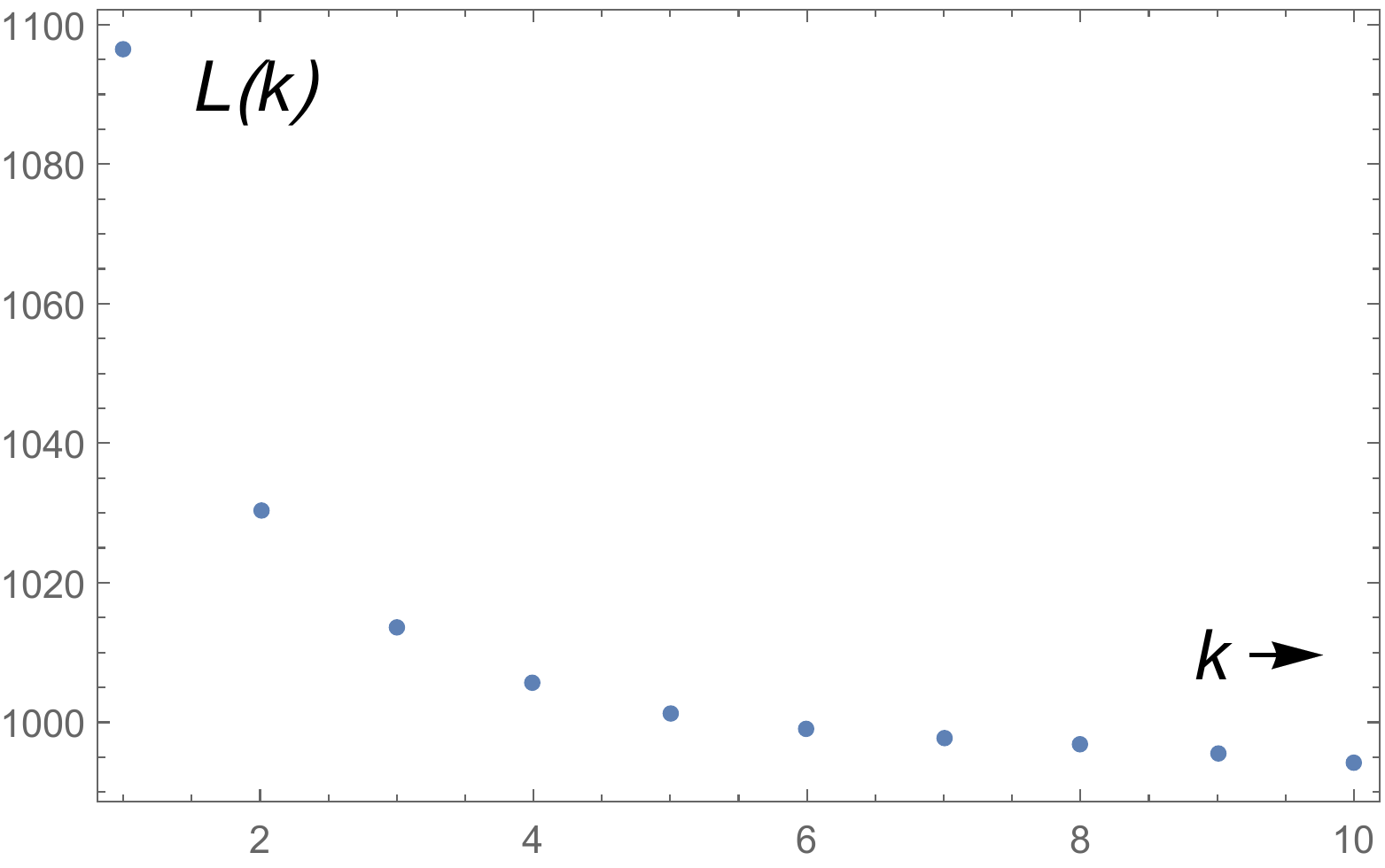}
\caption{Negative log-likelihood $L(k)$ as a function of the number of communities $k$. This plot is used to to find sufficiently optimal value of $k$.
The right value of $k$ is in the range of $6$ to $10$, because for larger values $L$ is approximately a constant.}
\label{cost}
\end{figure}

We studied a Gnutella peer-to-peer network \cite{gnutella} with $10876$ nodes representing hosts and $39994$ directed links representing connections between the hosts. The graph is sparse because the link density is just about $3.4\times 10^{-4}$. We extracted the largest strongly connected component which contains $n=4317$ nodes and ran the regular decomposition algorithm for the corresponding full distance matrix ($m=n$) for different values for the number of communities in the range $k=1,2,\dots,10$. From the corresponding plot (Fig.~\ref{cost}) of the negative log-likelihood function we decided that $k=10$ is valid choice.  

%
%
%
%

\begin{figure}
\centering
\includegraphics[width=8cm]{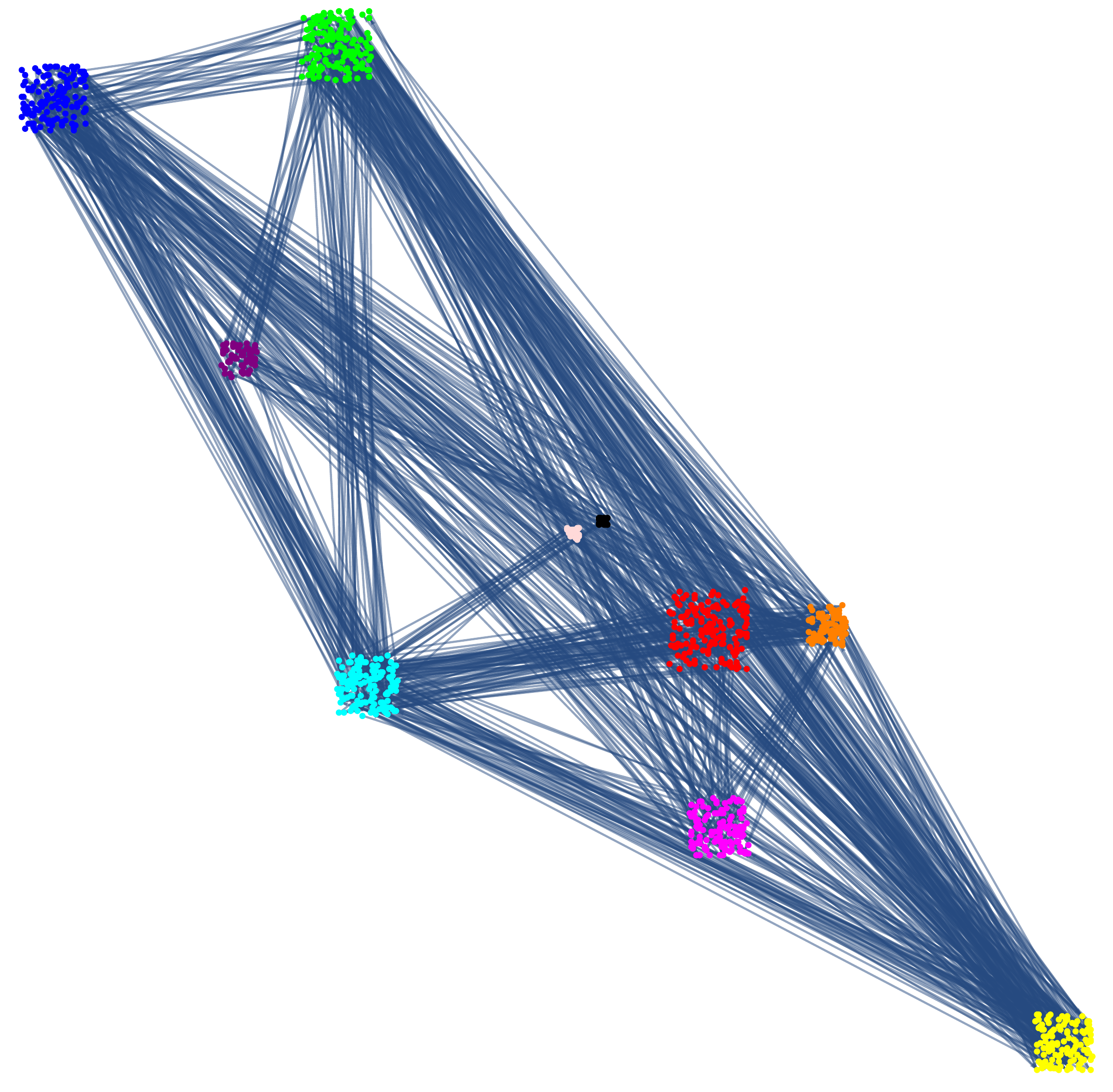}
\caption{The strongly connected component of a directed Gnutella network partitioned into 10 communities.}
\label{gnutella}
\end{figure}

\begin{figure}
\centering
\includegraphics[width=8cm]{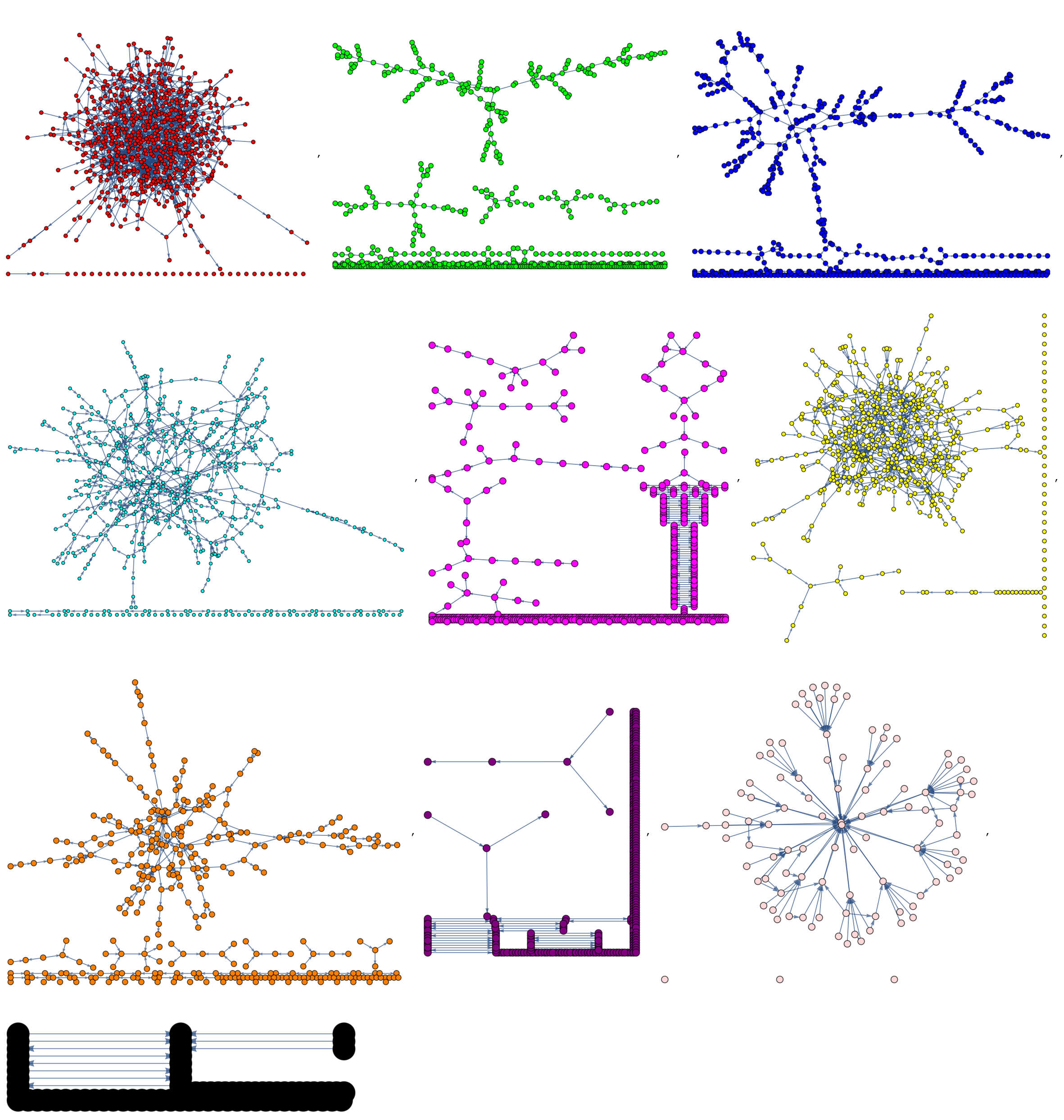}
\caption{Subgraphs induced by the 10 regular groups of the Gnutella network. The subgraphs are structurally significantly different from each other. For instance, the directed cycle counts of the subgraphs (ordered row by row from left to right) are 139045, 0, 0, 2, 0, 15, 3, 0, 0, 0. The first community might be identified as a core of the network. Corresponding sizes of regular groups can be seen in Fig.~\ref{gnutella} }
\label{gnutellagroups}
\end{figure}

Fig.~\ref{gnutella} illustrates the inter-community structure of the partitioned graph into $k = 10$ communities, and Fig.~\ref{gnutellagroups} describes the subgraphs induced by the communities. The induced subgraphs are internally quite different from each other. The high degree core-like parts form their own communities and they play a central role in forming paths through the network. Together these two figures provide a low-dimensional summary of the network as a weighted directed graph on $10$ nodes with self-loops and weights corresponding to link densities in both directions.


\subsection{Internet autonomous systems}

The next example is a topology graph of Internet's Autonomous Systems \cite{as} obtained from traceroute measurements, with around $1.7$ million nodes and $11$ million undirected links. This graph was analyzed using a HPC cluster.  

We used a simplified scheme to analyze this graph. This was dictated by limited time and also we wanted to test some heuristic ideas to speed-up regular decomposition even further. 
First we computed shortest paths between a hundred randomly selected pairs of nodes. Then $30$ most frequently appearing nodes in those shortest paths were selected as reference nodes. These nodes also appeared at the top of the link list provided by the source \cite{as}. That is why we assume that such an important ordering of nodes is used in this source data set. Next we took $2000$ top nodes from the source list and $3000$ uniformly random nodes from the set of all nodes. A distance matrix from the $m=30$ reference nodes to the selected $n=5000$ target nodes was computed. Then the regular decomposition algorithm was run on this distance matrix for different values of $k$. From the negative log-likelihood function plot an optimal number of communities was estimated to be $k=15$. As a result, we get a partition of the selected 5000 nodes into $15$ communities.

\begin{figure}
\centering
\includegraphics[width=9cm]{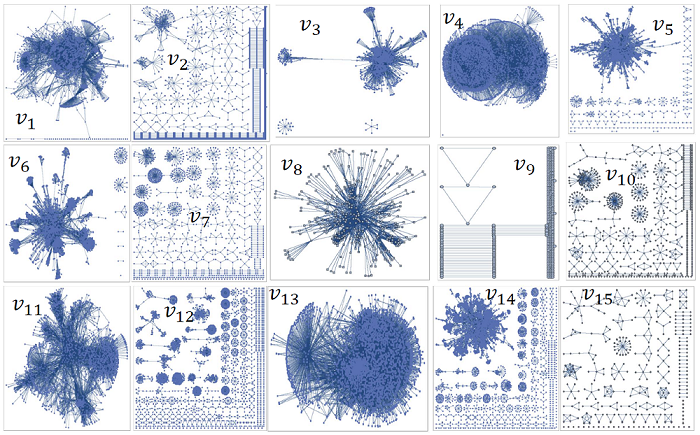}
\caption{\label{fig:as} Subgraphs of AS graph induced by 15 regular groups $v_1,\cdots, v_{15}$. The biggest $4$ subgraphs are represented by subgraphs induced by around 10 percent of the nodes of the group (these groups are $v_1, v_4, v_{11}, v_{13}$) and the rest of $11$ groups are fully depicted.}
\end{figure}

To enlarge the communities we used the following heuristic.  For each node belonging to one of the communities, we include all neighbors of the node to the same group. This can be  justified, since such neighbors should have very similar distance patterns as the root nodes. In this way a large proportion of nodes were included in the communities, more than 30 percent of all nodes. The result is partially shown in Fig.~\ref{fig:as}, some of the groups were very large, having around $3 \times 10^5$ nodes, and only part of them are plotted.

The found subgraphs are structurally heterogeneous and thus informative. For comparison, most subgraphs induced by a random samples of 1000 nodes contained no links in our experiments.

\section{Conclusions and future work}
This paper introduced a new approach for partitioning graphs using observed distances instead of usual path and cycle counts. By design, the algorithm easily scales to very large data sets, linear in the number of target nodes to be partitioned. First experiments presented here with real and synthetic data sets suggest that this method might be quite accurate, and possibly capable of reaching the Kesten--Stigum threshold. However, to be convinced about this, more detailed theoretical studies and more extensive numerical experiments are needed. We also need to estimate quantitatively accuracy of the low-dimensional approximation in synthetic cases like the random power-law graphs.  
%
%
Spectral methods utilizing the distance matrix as a basis of network analysis are of broader interest, see \cite{bollabook,hansen}. We are also interested in finding relations of our concept with graph limits in the case of sparse networks \cite{Borgs}, and extending the analytical result to sparse random graph models with nontrivial clustering \cite{Bloznelis_Leskela_2016,Karjalainen_Leskela_2017,Karjalainen_VanLeeuwaarden_Leskela_2018}. We aim to study stochastic block models with more than two groups and the actual distance distributions in such random graphs.

We will also find real-life applications for our method in machine learning such as highly topical multilabel classification, \cite{su,read,demb,bhatia,babbar }. For instance in case of natural language documents like news release, we can use deep-learning to embed words or paragraphs into points in a vector space. The Euclidean distance between corresponding vectors indicates affinity of meaning words etc. Our graph method could be used to analyze networks of large volumes of of such documents. Each document has usually more than one meaningful labeling. We will study possibilities of aiding such a multilabel classification using RD of the training data.


\section*{Acknowledgment}
This work was supported by ECSEL-MegaMaRT2 project.


\end{document}